\documentclass[a4paper]{article}
\usepackage{ISCSLP2022}
\usepackage{amsmath,graphicx}
\usepackage{xcolor}
\usepackage{amsthm, bm}
\usepackage{amssymb}
\usepackage{algorithmic}
\usepackage{array}
\usepackage{capt-of}
\usepackage{fixltx2e}
\usepackage{stfloats}
\usepackage{url}
\usepackage{framed,multirow,booktabs}
\usepackage{graphicx,epsfig}
\usepackage{amssymb,amsmath,bm}
\usepackage{textcomp}
\UseRawInputEncoding

\title{A Study on Joint Modeling and Data Augmentation of Multi-Modalities for Audio-Visual Scene Classification}
\name{Qing Wang\textsuperscript{1}, Jun Du\textsuperscript{1}, Siyuan Zheng\textsuperscript{1}, Yunqing Li\textsuperscript{1}, Yajian Wang\textsuperscript{1}, Yuzhong Wu\textsuperscript{2，4}, Hu Hu\textsuperscript{3},\\
Chao-Han Huck Yang\textsuperscript{3}, Sabato Marco Siniscalchi\textsuperscript{3，5}, Yannan Wang\textsuperscript{2}, Chin-Hui Lee\textsuperscript{3}}
\address{
  \textsuperscript{1}NELSLIP, University of Science and Technology of China, Hefei, China, \\
\textsuperscript{2}Tencent Ethereal Audio Lab, Tencent Corporation, China, \\
\textsuperscript{3}School of Electrical and Computer Engineering, Georgia Institute of Technology, GA, USA, \\
\textsuperscript{4}DSP \& Speech Technology Laboratory, The Chinese University of Hong Kong, Hong Kong, \\
\textsuperscript{5}Computer Engineering School,University of Enna Kore, Italy}
\email{jundu@ustc.edu.cn}

\begin{document}

\maketitle
\begin{abstract}
  In this paper, we propose two techniques, namely joint modeling and data augmentation, to improve system performances for audio-visual scene classification (AVSC). We employ pre-trained networks trained only on image data sets to extract video embedding; whereas for audio embedding models, we decide to train them from scratch. We explore different neural network architectures for joint modeling to effectively combine the video and audio modalities. Moreover, data augmentation strategies are investigated to increase audio-visual training set size. For the video modality the effectiveness of several operations in RandAugment is verified. An audio-video joint mixup scheme is proposed to further improve AVSC performances. Evaluated on the development set of TAU Urban Audio Visual Scenes 2021, our final system can achieve the best accuracy of 94.2\% among all single AVSC systems submitted to DCASE 2021 Task 1b. 
\end{abstract}

\section{Introduction}
\label{sec:intro}

Acoustic scene classification (ASC) task aims to identify the environment classes of audio recordings, which can be used for various demands of contextualization and personalization. The Detection and Classification of Acoustic Scenes
and Events (DCASE) Challenge has organized ASC related tasks for years \cite{mesaros2017dcase,mesaros2018multi,heittola2020acoustic} under different scenarios. In this year, DCASE 2021 Challenge proposes a new audio-visual scene classification (AVSC) task \cite{wang2021curated} by leveraging on additional information of video modality, which makes a big difference compared to previous tasks. Since multi-modal methods can greatly boost the performance compared to single modality \cite{wang2021curated}, AVSC should be more promising in challenging realistic applications.
%The key is to jointly model audio and visual modalities.

For an ASC task, the most competitive methods \cite{battaglino2016acoustic,hu2021two} in the previous DCASE Challenges extracted representative audio embedding by inputting log-Mel spectrogram or Mel-frequency cepstral coefficients (MFCC) features into deep neural networks for classification. Data augmentation strategies such as mixup \cite{zhang2017mixup} and SpecAugment \cite{park2019specaugment} were also introduced to enhance generalization ability of models. For visual scene classification (VSC), many convolutional neural networks (CNNs) based methods  \cite{simonyan2014very,he2016deep, huang2017densely} transferred from object recognition task were adopted.  Large in-domain datasets \cite{xiao2010sun, zhou2017places} were also utilized for pre-training, which could enrich the extracted feature. Additionally, there exist other group of networks \cite{hayat2016spatial, liu2018dictionary} which were specifically developed for scene classification.
%In the AVSC task, the addition of visual information brings new challenges: (1) the jointly modeling of two modalities. (2) the joint data augmentation strategy.
For the new AVSC task \cite{wang2021curated}, audio recordings and corresponding video clips are both provided, which means new model architectures and new training strategies are needed. Multi-modal fusion approaches can be summarized into three categories: early fusion \cite{castellano2008emotion}, late fusion \cite{ramirez2011modeling} and hybrid fusion \cite{lan2014multimedia}. Since early fusion strategy can learn the correlation of different modalities at the feature level, it has become a very popular way to tackle multi-modal fusion.

In this paper, we propose a simple yet effective multi-modal approach for AVSC task, which mainly consists of audio module, visual module and modality fusion module.
Under the premise of multi-modal input, we explore the selection of unimodal representation to improve system performance.
Our observation is that the combination of audio feature extraction with our customized fully convolutional neural networks (FCNN) and video feature extraction with DenseNet \cite{huang2017densely} can achieve the best results.
The paradigm of joint fine-tuning after pre-training is also introduced in AVSC task, which can greatly enhance the performance in Task 1b of DCASE 2021 Challenge.
Moreover, we design multi-modal data augmentation strategies that fully enrich the input diversity of two modalities.
In particular, a joint mixup strategy is proposed to synchronously generate new audio and video data, thus adding modality correlation at the input level.
Evaluated on DCASE 2021 Task 1b, experimental results show that our system achieves the best accuracy of 94.2\% on the development set among all single systems.

\section{The proposed AVSC approach}
\label{sec:Audio-visual Scene Classification}

\subsection{Acoustic-Visual Joint Modeling}
\begin{figure}[h]
	\centering
	\includegraphics[width=3.1in]{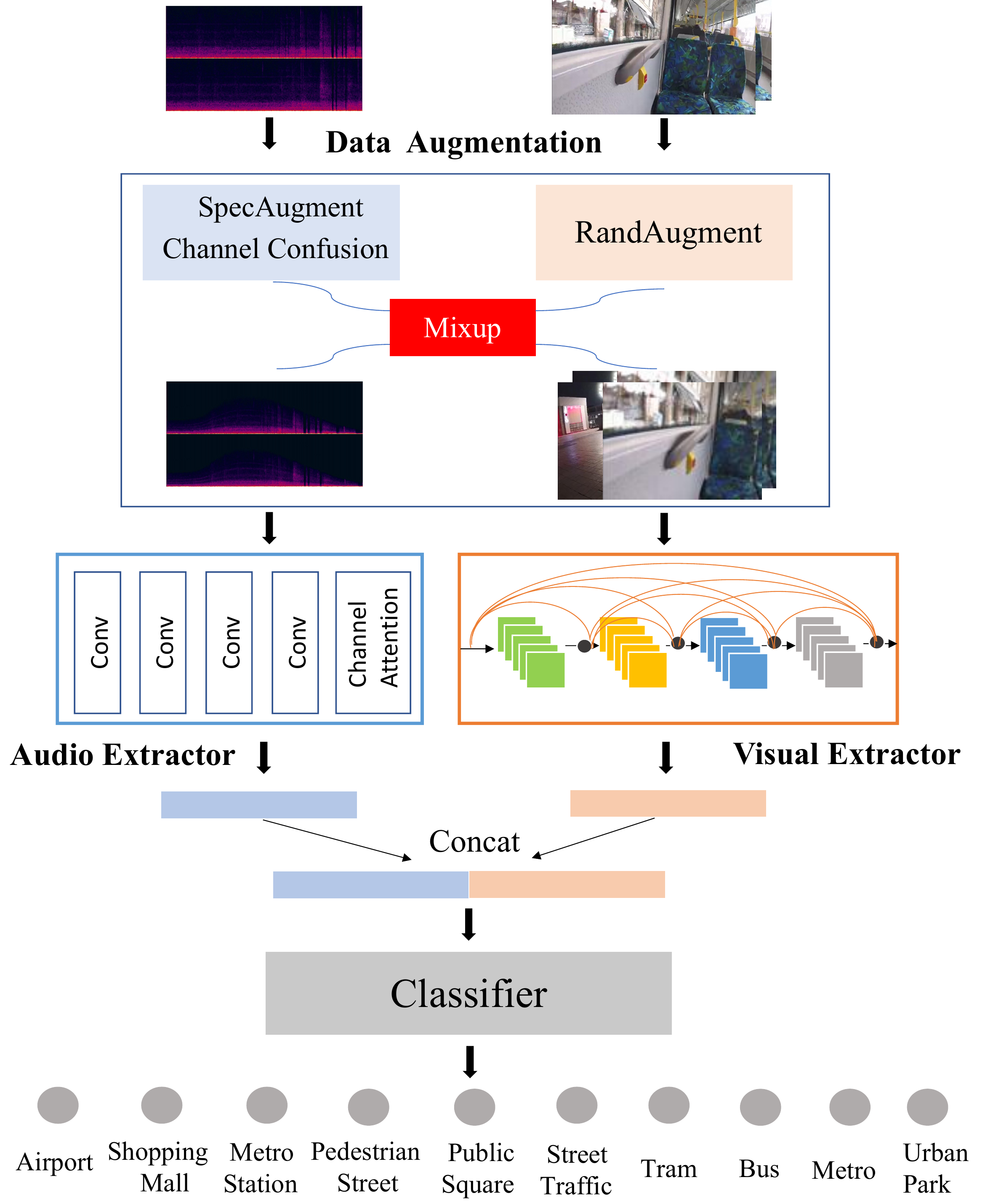}
	\caption{The proposed AVSC system.}
	\label{fig1}
\end{figure}

For the AVSC task, we design a multi-modal system as shown in Fig.\ref{fig1}.
Our AVSC model mainly consists of three parts: audio module, visual module and modality fusion module. We will elaborate them in the following subsections.

\subsubsection{Audio Embedding}
In the audio module,
%\textcolor{red}{the input to the network consists in a multi-channel 3D audio representation compiling information from the left and right input audio channels as well as their difference.}
we extract the log-Mel filter bank (LMFB)
features of the raw data $\bm{I}_\text{A}$ with delta and delta-delta operations, forming the input $\bm{I}_\text{A}^\text{LMFB}$.
For high-level feature representations, we employ audio extractor $f_\text{A}$ on $\bm{I}_\text{A}^\text{LMFB}$, and obtain the audio embedding $\bm{E}_\text{A}$ as:
\begin{equation}
    \bm{E}_\text{A} = f_\text{A}(\bm{I}_\text{A}^\text{LMFB})
\end{equation}

FCNN model is applied to extract high-level audio embedding, which achieved promising performance for the ACS task in our previous work \cite{hu2021two}.
The FCNN model mainly consists of four convolution blocks with total 9 stacked convolutional layers and a channel attention block. Each convolutional layer is followed by a batch normalization and ReLU activation function. Dropout is used to alleviate over-fitting. Channel attention \cite{bi2020multiple} is applied before the final global average pooling layer. In addition to the FCNN model which is trained from scratch, we also investigate an audio feature extractor VGGish \cite{hershey2017cnn} pre-trained on AudioSet \cite{gemmeke2017audio}.
% \textcolor{red}{The dimension of audio embedding vector obtained from FCNN is 128.}

\subsubsection{Video Embedding}
In the visual module, high-level video embedding $\bm{E}_\text{V}$ is calculated given the input image $\bm{I_\text{V}}$ by using visual extractor $f_V$ as follows,
\begin{equation}
    \bm{E}_\text{V} = f_V(\bm{I_\text{V}}).
\end{equation}

We explore the effect of different networks (discussed in Section \ref{experiment_joint_modeling}), and select DenseNet \cite{huang2017densely} as our video embedding extractor.
% ResNeSt proposed Split-Attention Block by combining the channel-wise attention with multi-path network for more diverse representations.
DenseNet proposed dense connection between layers  in a feed-forward fashion, which encouraged the maximum information flow in network.
The transfer learning results on many downstream tasks, such as object detection and instance segmentation have proved its effectiveness.
To enhance the generalization ability of the model, we adopt models pre-trained on in-domain large databases and apply them in this AVSC task by transfer learning.
% \textcolor{red}{After extractor, the dimension of visual embedding from DenseNet is 1280.}

\subsubsection{Modality Fusion}
With the high-level feature embedding of audio and video, modality fusion module is needed to integrate information from two sources.
To fully exploit the complementarity of two modalities, we concatenate audio embedding $\bm{E}_\text{A}$ and video embedding $\bm{E}_\text{V}$ into fusion embedding $\bm{E}_\text{F}$, and feed it into a classifier, which is a four-layer multi-layer perceptron (MLP) network with size of 512, 128, 64 and 10, respectively.
\begin{equation}
    \bm{E}_\text{F} = \left[ \bm{E}_\text{A}, \bm{E}_\text{V} \right]
\end{equation}
\begin{equation}
    \bm{p} = \text{MLP}(\bm{E}_\text{F})
\end{equation}
The output probability vector $\bm{p}=[p_1, p_2, ..., p_{10}]$ is a 10-dimensional vector, corresponding to the number of classes, i.e. airport, shopping mall, metro station, pedestrian street, public square, street traffic, tram, bus, metro and urban park.

Suppose that the scene label is one-hot vector $\bm{y}=[y_1, y_2, ..., y_{10}]  \in \mathbb{R}^{10}$, the cross entropy loss for classification is calculated as follows:
\begin{equation}
    L = - \sum_{i=1}^{10} ~y_i~\text{log}~p_i
\end{equation}

It is worth noting that we train the whole AVSC model without freezing the audio or visual extractor, which means the parameters of both extractors are updated together with the classifier parameters during model training.  Unlike feature-based approaches \cite{wang2021audio, okazaki2021ldslvision}, our fine-tuning strategy makes two feature extractors more task-specific on the challenge data set. Moreover, fine-tuning the two modalities together can achieve better modality fusion performances.

\subsection{Audio-Video Data Augmentation}
  In our previous work \cite{hu2021two}, data augmentation methods are proven to be effective in ASC task of DCASE 2020 Challenge. Thus in this task, we adopt several different data augmentation methods for audio and video modalities. In addition, a joint mixup strategy is proposed, which is effective to improve the generalization abilities of models.

\subsubsection{Audio Data Augmentation}
  When training the audio extractor, four techniques that generate extra data are used as listed below: (i) pitch shifting, where we perform it on audio clips to randomly shift the pitch based on the uniform distribution; (ii) speed changing, where we perform it on audio waveforms to randomly change the speed of audio recordings; (iii) noise adding, where we perform it on audio waveforms to add random Gaussian noise; and (iv) audio mixing, where we mix two audio recordings from the same scene class  to generate a new sample with the same label. The other two on-the-fly data perturbation techniques includes: (i) SpecAugment \cite{park2019specaugment}: in this study, we do not perform time warping. For time and frequency masking, we set the masking parameter to 10 \% of the dimensions. It is applied on LMFB features and is performed on batch level; (ii) channel confusion: two channels of input audio feature are randomly swapped.
  % and (iii) mixup: It was proposed in \cite{zhang2017mixup} and the parameter alpha is set to 0.4 in our work. Two batches of data are randomly mixed in each step along with the corresponding labels when training audio models.
%   After training the audio extractor，three on-the-fly data augmentation schemes are applied to audio model fine-tuning for
  The two on-the-fly audio data perturbation strategies are also applied when fine-tuning audio extractor.

\subsubsection{Video Data Perturbation}
%   To train the visual extractor, we use DenseNet161  pretrained on Places365 database proposed for scene recognition.
  In order to improve the robustness of the visual extractor, we investigate data augmentation techniques in RandAugment \cite{9150790}, which was first used for image object classification and object detection tasks. RandAugment is a search algorithm to find the best data augmentation strategy. There are two optimal parameters called $N$ and $M$. For each image in each mini-batch, select $N$ data augmentation methods with $M$ magnitudes.
  Due to the difference between the object image and  scene image, the accuracy drops after directly applying RandAugment to the AVSC task. Therefore, we evaluate each sub-policy for its usefulness in scene classification. Through detailed ablation studies, three sub-policies in the search space, namely Sharpness, Contrast and Identity Mapping are adopted. For each image, two operations are randomly selected to be applied in sequence to image data. Since these video augmentation techniques do not produce any extra offline data, we also call it data perturbation to make a distinction from commonly used data augmentation strategies. It is worth noting that all the video data perturbation techniques mentioned above are adopted for fine-tuning visual extractor. 

\subsubsection{Audio-Video Joint Mixup}
%   After training the audio extractor and visual extractors, we conduct audio-visual joint modeling for AVSC. We follow previously-proposed data augmentation strategies. Three on-the-fly data augmentation schemes are applied to audio model fine-tuning, while video date perturbation mentioned above is adopted for fine-tuning visual models. 
  Mixup was firstly proposed for improving the robustness of deep neural networks, which has been successfully applied in various tasks, such as unimodal image classification and adversarial examples generating.
  However, in multi-modal tasks,  mixup has not been investigated yet as far as we know. 
  With the change of input, regular mixup strategy cannot be directly applied in our AVSC task, let alone the verification of its effectiveness.

  Here in this section, we propose a novel joint mixup algorithm, successfully applying the mixup to the audio and visual inputs at the same time. 
  Our joint mixup method mixes the input of two modalities synchronously, further increasing the diversity of input for better enhancing our joint AVSC model.
  In more detail, a joint audio-video example can be constructed by using the following formula:
\begin{equation}
    (x^\text{ma}_{ij}, x^\text{mv}_{ij}) = \alpha \times(x^\text{a}_{i},x^\text{v}_{i})+(1-\alpha )\times(x^\text{a}_{j}, x^\text{v}_{j})
\end{equation}
\begin{equation}
    t^\text{m}_{ij} = \alpha \times t_{i} +(1-\alpha )\times t_{j} \ \ \ \ \ \ \
\end{equation}
where $ (x^\text{a}_{i},x^\text{v}_{i}) $ are audio and video  embedding of the $i$-th sample and $ t_{i} $ is the corresponding label.
$x^\text{ma}$, $x^\text{mv}$ and $t^\text{m}$ denote the audio, video embedding and scene label after the joint mixup, respectively.
$\alpha$ is the mixed ratio and set to 0.4.
%the same as used in the audio pre-trained model. As a reason, it  develops  non-zero values at two positions on the label embedding. 
To build the multi-modal systems, 20\% of the training data are employed with audio-video joint mixup.

\section{Experimental Results and Analysis}
\label{sec:experiment}

\subsection{Experimental Setup}
The data set used for the AVSC task is TAU Urban Audio-Visual Scenes 2021 \cite{wang2021curated}, which consists of 34 hours of data with time-synchronized audio and video content. There are about 8k 10-second audio clips for training and 3k test audio clips recorded in a binaural way using a 48kHz sampling rate. For the video clips, every second contains 30 frames. We split all the data into 1-second samples without overlap to match the development set of Task 1b in DCASE 2021 Challenge.

For each input audio clip, we use the Librosa \cite{2015librosa} library to extract the LMFB features and compute log-Mel delta and delta-delta operations without padding, which generates a feature tensor shape of $39\times128\times6$. For each input visual clip, the first frame and the fifteenth frame images are extracted and resized to $224\times224$ patches to calculate the video embedding. Then two video embedding are added together to serve as the final visual feature of the input video data. All our models are trained using PyTorch toolkit. And Adam optimizer is used during training. For our audio-visual joint model, we set a small learning rate of 1e-5 to fine-tune both audio and video extractors, with a weight decay of 1e-5 and batch size of 32.

\subsection{Results on Acoustic-Visual Joint Modeling}
\label{experiment_joint_modeling}
\begin{table}[]
	\centering
	\footnotesize
	\caption{An accuracy (`Acc.' in \%) comparison of acoustic-visual joint modeling. The first two columns correspond to audio and video models, where `VGGish' is pre-trained on AudioSet while `FCNN' is trained from scratch. `Pre\_V' denotes pre-trained video models. `Fine\_A' and `Fine\_V' denote fine-tuning  the audio and video extractors, respectively.}
	\vspace{5pt}
	\label{tab1}
	\begin{tabular}{c|c|c|c|c|c}
		\toprule
		Audio    & Video       & Pre\_V & Fine\_A & Fine\_V & Acc. \\ \toprule
		VGGish \cite{hershey2017cnn} &- &- &\checkmark &-  &59.3 \\
		FCNN &- &- &- &- &75.2\\
		- &VGG &- &- &- &76.2\\
		- &VGG &\checkmark &- &- &80.3\\
		%ResNet50 & ResNet18    & -   & -       & -       &   75.0     \\
		FCNN & VGG  &\checkmark   & -       & -       &   87.4     \\
		%FCNN     & ResNet18    & -   & -       & -       &   80.4       \\
		
		%FCNN     & ResNet18    & \checkmark   & -       & -       &   85.2       \\
		FCNN     & ResNet    & \checkmark   & -       & -       &   91.6      \\
		FCNN     & ResNeSt   & \checkmark   & -       & -       &   92.2       \\
		FCNN     & DenseNet & \checkmark   & -       & -       &   92.5       \\ \toprule
		FCNN     & DenseNet & \checkmark   & \checkmark       & -       &      93.1    \\
		FCNN     & DenseNet & \checkmark   & -       & \checkmark       &    92.9      \\
		FCNN     & DenseNet & \checkmark   & \checkmark       & \checkmark      &     93.2     \\ \toprule
	\end{tabular}
\end{table}
We consider three aspects for acoustic-visual joint modeling: the selection of audio extractor, the selection of visual extractor, and the joint training strategy. Here in this section, we conduct a series of experiments to show the effectiveness of our joint modeling. Table \ref{tab1} shows the experimental results on the development set.

For audio embedding, we compare VGGish \cite{hershey2017cnn} which is pre-trained on AudioSet and FCNN which is trained from scratch. All these preliminary experiments are conducted  without data augmentation methods.
From the top two rows of Table \ref{tab1}, we can see that the FCNN model trained from scratch performs better than pre-trained VGGish.
%as the audio extractor greatly exceeds ResNet50 by an absolute accuracy gain of 5.4\%.
% Please note that VGGish and PANN are pretrained on the AudioSet, while FCNN and ResNet are trained from scratch on the official data. From Table \ref{tab2}, we can  see that models without pretraining uniformly perform better with the gap of about 3\%.
% Additionally, FCNN exceeds ResNet50 by \textcolor{red}{xx\%}.
Therefore, in the following experiments, FCNN trained from scratch with the official data is used to extract audio embedding.

For visual embedding, large in-domain data sets, such as ImageNet \cite{deng2009imagenet} are always adopted for pre-training. Firstly, we make a comparison of video models with and without pre-training. From the third and fourth rows of Table \ref{tab1}, VGG model pre-trained on ImageNet achieves higher accuracy than that trained from scratch, which demonstrates that pre-training is helpful for extracting better visual embedding. In this study, we adopt another scene-image data set Places365 \cite{zhou2017places} for pre-training and
compare different visual embedding. ResNet, ResNeSt and DenseNet are all pre-trained on the Places365 data set.
%As shown in Table \ref{tab1}, a comparison between the fifth to eighth rows and the third row reflects that pre-training is helpful for extracting better visual features with an absolute improvement of 4.8\%.
% The same model ResNet50 pretrained on different datasets have different performance: V4 outperforms V3 by 0.3\%, which shows scene-image classification dataset Places365 works more suitable on our AVSC task.
We compare the performance of four outstanding pre-trained networks for video feature extracting as shown between the fifth to eighth rows of Table \ref{tab1}. The results show that, in the AVSC model, FCNN for audio extractor and DenseNet for visual extractor achieve the best performance with an accuracy of 92.5\%.

\begin{table}[!htp]
\centering
\caption{An accuracy (in \%) comparison on the DCASE development set of different video data perturbation strategies to modify each individual video feature.}
\vspace{5pt}
\label{tab2}
\begin{tabular}{c|c|c}
\hline
Index & Transformation  & Accuracy  \\ \hline
1     & A-FCNN+V-DenseNet & 93.2                \\ \hline
2     & TranslateX      & 92.6               \\ \hline
3     & TranslateY      & 92.6                 \\ \hline
4     & Solarize        & 90.4                 \\ \hline
5     & ShearX          & 91.5                 \\ \hline
6     & ShearY          & 91.0              \\ \hline
7     & Sharpness       & 93.3                \\ \hline
8     & Rotate          & 92.0                 \\ \hline
9     & Posterize      & 88.5         \\ \hline
10    & Invert         & 91.3         \\ \hline
11    & Equalize       & 92.7         \\ \hline
12    & Cutout         & 91.5         \\ \hline
13    & Contrast       & 93.5         \\ \hline
14    & Color          & 92.3         \\ \hline
15    & Brightness     & 92.0         \\ \hline
16    & AutoContrast   & 93.2         \\ \hline
\end{tabular}
\end{table}

For joint training, the extractor for each modality has two choices: freezing  or fine-tuning the parameters. Based on the best extractors (FCNN and DenseNet), we compare three different joint training strategies shown in the bottom three rows of Table \ref{tab1}. We can conclude that fine-tuning can greatly improve performances in audio-visual models. Fine-tuning only the audio or visual extractor achieves better results than no fine-tuning. Fine-tuning both audio and visual extractors can achieve the best performance of 93.2\%. That is what we adopt for our follow-up experiments.

\subsection{Results on Audio-Video Data Augmentation}
We have shown that data augmentation is effective for the ASC task in \cite{hu2021two}. On the other hand, not all  data augmentation methods are valid for the VSC task. Accordingly, we now investigate the effectiveness of the 15 data perturbation strategies in RandAugment. Table \ref{tab2} shows the accuracy for 15 video data perturbation policies of the joint audio-visual model. The audio modality in the joint model is trained without any data augmentation. There are two parameters, the number of data augmentation methods $N$ and the magnitude $M$. 
%We set the data augmentation probability to be 50\%, which means either `Identity' or `using data augmentation method'. 
The parameters $N$ and $M$ are set to 2 and 14, respectively. The top row in Table \ref{tab2} provides the accuracy of the joint system without any data augmentation. By comparing the results in Table \ref{tab2}, we can observe that `Sharpness' and `Contrast' are effective to improve system performances for scene classification, while other methods in RandAugment are not suitable for this task. Eventually, we adopt `Sharpness', `Contrast' and `Identity Mapping' as the sub-policies of RandAugment.

The experimental results of audio-visual joint model with different data augmentation methods are presented in Table \ref{tab3}. Clearly, both audio and video data augmentation methods are effective, while the use of both can further improve the performance. Specifically, the accuracy when adopting both audio and video data augmentation is improved to 93.9\%.
We apply the mixup method for the audio-visual joint model. It has been proved effective for the audio system. Nevertheless, there is no idea about what proportion of mixing can bring the greatest improvement for video clips. Accordingly, the performance comparisons of video model when using different mixup percentage are listed in Table 3. The best accuracy of 94.2\% can be achieved when doing mixup on 20\% of the data, while the accuracy of audio-visual joint model baseline is 93.2\%. Consequently, we set the joint mixup percentage of 20\% for the final audio-visual model based on the results of Table 3.

\begin{table}[]
	\centering
	\caption{Experimental results of accuracy (in \%) for joint audio-visual data augmentation. The middle three columns denote whether audio augmentation, video perturbation and joint mixup (in term of percentage of training data) are applied. `\checkmark' and `-' denote with and without each operation, respectively.}
	\vspace{5pt}
	\label{tab3}
	\begin{tabular}{c|c|c|c|c}
		\hline
		System                                                                             & Audio & Video & Mixup & Accuracy \\ \hline
		\multirow{7}{*}{\begin{tabular}[c]{@{}c@{}}A-FCNN\\ +\\ V-DenseNet\end{tabular}} & - & - & -     &   93.2       \\
		& \checkmark & - & -     &     93.4     \\
		& - & \checkmark & -     &    93.7      \\
		& \checkmark & \checkmark & -     &    93.9      \\ \cline{2-5}
		& \checkmark & \checkmark & 20\%   &    94.2      \\
		& \checkmark & \checkmark & 50\%   &    93.8     \\
		& \checkmark & \checkmark & 100\%   &    94.0      \\ \hline
	\end{tabular}
\end{table}

\begin{table}[]
\centering
\caption{An accuracy (in \%) comparison of the state-of-the-art techniques.}
\vspace{5pt}
\label{tab4}
\begin{tabular}{c|c|c}
\hline
System & Model              & Accuracy \\ \hline
1     & Baseline \cite{wang2021curated}     & 77.0         \\ \hline
2     & Zhang et al. \cite{wang2021audio} & 94.1         \\
3     & Yang et al. \cite{Yang2021}  & 93.9         \\
4     & Pham et al. \cite{Pham2021}  & 93.9         \\ \hline
5     & Proposed  & 94.2         \\ \hline
\end{tabular}
\end{table}

\subsection{Overall Comparison}
The top row in Table 4 lists the AVSC accuracy of the official baseline system. Based on the original OpenL3 publication, either audio and video embedding is extracted using a single modality. Then it connects the sub-networks using two fully-connected feed-forward layers. Compared with the baseline system, our proposed model achieves promising performance by a large margin. Furthermore, we compare the system results of other teams with the top rankings for DCASE 2021 Task 1b. They are shown in the 2nd, 3rd and 4th rows in Table 4. The accuracy of our final AVSC system is listed in the bottom row. The experimental results clearly demonstrate that the proposed audio-visual joint model with audio-video data augmentation achieves quite a competitive performance when compared with all the  state-of-the-art benchmark systems.

\section{Conclusion}
\label{sec:conc}
In this paper, we propose a novel approach to the AVSC task with acoustic-visual joint modeling and data augmentation strategies. Based on multi-modal inputs, we compare different audio and video embedding and achieve the best matching result: FCNN for the audio modality and DenseNet for the video modality. Joint modeling with fine-tuning on both modalities works best with good accuracy improvements. Moreover, we successfully apply RandAugment with `Sharpness' and `Contrast' policies in training AVSC models.
% and 20\% data with mixup get a maximum improvement.
A joint mixup strategy for multi-modal AVSC is then proposed for better modality interactions at the input level. Finally, our system obtains the state-of-the-art performance with an accuracy of 94.2\% on the development set of DCASE 2021 Task 1b.

\newpage
\bibliographystyle{IEEEtran}

\bibliography{mybib}

% \begin{thebibliography}{9}
% \bibitem[1]{Davis80-COP}
%   S.\ B.\ Davis and P.\ Mermelstein,
%   ``Comparison of parametric representation for monosyllabic word recognition in continuously spoken sentences,''
%   \textit{IEEE Transactions on Acoustics, Speech and Signal Processing}, vol.~28, no.~4, pp.~357--366, 1980.
% \bibitem[2]{Rabiner89-ATO}
%   L.\ R.\ Rabiner,
%   ``A tutorial on hidden Markov models and selected applications in speech recognition,''
%   \textit{Proceedings of the IEEE}, vol.~77, no.~2, pp.~257-286, 1989.
% \bibitem[3]{Hastie09-TEO}
%   T.\ Hastie, R.\ Tibshirani, and J.\ Friedman,
%   \textit{The Elements of Statistical Learning -- Data Mining, Inference, and Prediction}.
%   New York: Springer, 2009.
% \bibitem[4]{YourName17-XXX}
%   F.\ Lastname1, F.\ Lastname2, and F.\ Lastname3,
%   ``Title of your INTERSPEECH 2022 publication,''
%   in \textit{Interspeech 2022 -- 23\textsuperscript{rd} Annual Conference of the International Speech Communication Association, September 18-22, Incheon, Korea, Proceedings, Proceedings}, 2022, pp.~100--104.
% \end{thebibliography}

\end{document}